%% file: main.tex
\documentclass{sig-alternate-10pt}
\usepackage{listing}
\usepackage{eepic}
\usepackage{epic}
\usepackage{xspace}    
\usepackage{graphicx}

\usepackage{latexsym}
\usepackage[footnotesize]{subfigure}
\usepackage{afterpage}
\usepackage{amstext}   
\usepackage{url}
\usepackage{alltt}
\usepackage{booktabs}
\usepackage{epstopdf}
\usepackage{xcolor}
\usepackage{color}
\usepackage{morefloats}
\usepackage{footmisc}
\usepackage{enumitem}
\usepackage{amsmath}
\usepackage[normalem]{ulem}
\usepackage[absolute]{textpos}

\definecolor{heraldBlue}{rgb}{0.0,0.0,0.8}
\definecolor{heraldRed}{rgb}{0.8,0.0,0.0}
\definecolor{heraldGray}{rgb}{0.4,0.4,0.4}
\definecolor{heraldBlack}{rgb}{0.0,0.0,0.0} 
\definecolor{heraldGreen}{rgb}{0.0,0.4,0.0} 

\mathcode`\*="2201 

\newcommand{\ie}{{\em i.e.},\xspace}
\newcommand{\eg}{{\em e.g.},\xspace}

\newcommand{\bi}{\begin{itemize}}
\newcommand{\ei}{\end{itemize}}
\newcommand{\be}{\begin{enumerate}}
\newcommand{\ee}{\end{enumerate}}
\newcommand{\bc}{\begin{center}}
\newcommand{\ec}{\end{center}}

\DeclareRobustCommand{\eg}{e.g.\@\xspace} 
\DeclareRobustCommand{\ie}{i.e.\@\xspace}
\makeatletter
\DeclareRobustCommand{\etc}{%
    \@ifnextchar{.}%
        {etc}%
        {etc.\@\xspace}
}
\makeatother

\begin{document}


\title{I Always Feel Like Somebody's Watching Me \\ \huge{\it  Measuring Online
Behavioural Advertising}\\
\vspace{5mm}
\normalsize [To appear in ACM CoNEXT 2015 (please cite the conference version of this paper)]}

\numberofauthors{3}
\author{
\alignauthor Juan Miguel Carrascosa  \\
\affaddr{Univ. Carlos III de Madrid} \\
\email{\small jcarrasc@it.uc3m.es}
\alignauthor Jakub Mikians \\
\affaddr{Universitat Polytechnic Catalunya} \\
\email{\small jakub.mikians@gmail.com}
\alignauthor Ruben Cuevas\\
\affaddr{Univ. Carlos III de Madrid} \\
\email{\small rcuevas@it.uc3m.es}
\and
\alignauthor Vijay Erramilli \\
\affaddr{Guavus} \\
\email{\small evijay@gmail.com}
\alignauthor Nikolaos Laoutaris\\
\affaddr{Telefonica Research} \\
\email{\small nikolaos.laoutaris@telefonica.com}
}

\maketitle
\begin{abstract}

Online Behavioural \emph{targeted} Advertising (OBA) has risen in prominence as a method to increase the effectiveness of online advertising. OBA operates by associating tags or labels to users based on their online activity and then using these labels to target them. This rise has been accompanied by privacy concerns from researchers, regulators and the press. In this paper, we present a novel methodology for measuring and understanding OBA in the online advertising market. We rely on training artificial online \emph{personas} representing behavioural traits like `cooking', `movies', `motor sports', etc. and build a measurement system that is automated, scalable and supports testing of multiple configurations. 
We observe that OBA is a frequent practice and notice that categories valued more by advertisers are more intensely targeted. In addition, we provide evidences showing that the advertising market targets sensitive topics (e.g, religion or health) despite the existence of regulation that bans such practices. We also compare the volume of OBA advertising for our personas in two different geographical locations (US and Spain) and see little geographic bias in terms of intensity of OBA targeting. Finally, we check
for targeting with do-not-track (DNT) enabled and discovered that DNT is not yet enforced in the web.
\end{abstract}


\input{introduction}

\input{methodology_v2}

\input{system}

\input{results3}
\input{related_work}

\input{conclusion}


%
\begin{footnotesize}
\bibliographystyle{plain}
\bibliography{references}
\end{footnotesize}


\end{document}

%% file: introduction.tex

\section{Introduction}
\label{sec:intro}

Business models around \emph{personal information}, that include monetizing personal information via  Internet advertising and e-commerce~\cite{Mikians12:hotnets}, are behind most free Web services. 
Information about consumers browsing for products 
and services is collected, \eg, using tracking cookies, for the purpose of developing tailored advertising and e-marketing 
offerings (coupons, promotions, recommendations, \etc). While this can be beneficial for driving web innovation, companies, 
and consumers alike, it also raises several concerns around its privacy implications. There is a fine line between what 
consumers value and would like to use, and what they consider to be overly intrusive. Crossing this line can induce users to 
employ blocking software for cookies and advertisements \cite{ADBLOCKPLUS,disconnect_me,GHOSTERY,PRIVACYCHOICE}, 
or lead to strict regulatory interventions. Indeed, this economics around personal information has all the characteristics 
of a ``Tragedy of the Commons'' (see Hardin~\cite{Hardin68:Tragedy}) in which consumer privacy and trust towards the web and 
its business models is a shared \emph{commons} that can be over-harvested to the point of destruction. 

The discussion about privacy red-lines has just started\footnote{FTC released in May 2014 a report entitled ``Data Brokers -- A Call 
for Transparency and Accountability'', whereas the same year the US Senate passed the ``The Data Broker Accountability and 
Transparency Act (DATA Act)''.} and is not expected to conclude any time soon. Still, certain tactics, have 
already gained a taboo status from consumers and regulators, \eg, price discrimination in e-commerce~\cite{Mikians12:hotnets,Mikians13:conext,hannak-2014-ecommerce, Odlyzko2014:CapitalismDestruction}. Online Behavioural \emph{targeted} Advertising (OBA, Sec.~\ref{sec:behavioural}) on sensitive categories like sexual orientation, health, political beliefs etc.~\cite{Sweeney:2013}, or tricks to evade privacy protection mechanisms, like Do-Not-Track signals, are additional tactics that border the tolerance of most users and regulators.  The objective of this work is to build a reliable methodology for detecting, quantifying and characterizing OBA in display advertising on the Web and then use it to check for controversial practices.
\vspace{2pt}

\noindent {\bf Challenges in detecting Online Behavioural Advertising:}  The technologies and the ecosystem for delivering targeted advertising is truly mind boggling, involving different types of entities, including Aggregators, Data Brokers, Ad Exchanges, Ad Networks, \etc, that might conduct a series of complex online auctions to select the advertisement that a user gets to see upon landing on a webpage (see~\cite{Shuai:AdvertisingSurvey} for a tutorial and survey of 
relevant technologies). Furthermore, targeting can be driven by other aspects \eg, location, gender, age group, that have nothing to do with specific behavioural traits that users deem as sensitive in terms of privacy, or it can be due to ``re-targetting''~\cite{farahat2011retargeting,helft2010retargeting,lambrecht2013does} from previously visited sites.  
Distinguishing between the different types of advertising is a major challenge towards developing a robust detection technique for OBA. On a yet deeper level, behaviours, interests/types, and relevant metrics have no obvious or unique definition that can be used 
for practical detection. It is non-trivial to unearth the relative importance of different interests or characteristics that can be used 
for targeting purposes or even define them. 
Last, even if definitional issues were resolved, how would one obtain the necessary datasets and 
automate the process of detecting OBA at scale?

\vspace{2pt}

\noindent {\bf Our contribution:} The main contribution of our work is the development of an extensive methodology for detecting and characterizing OBA at scale, which allows us to answer essential questions: $(i)$ How frequently is OBA used in online advertising?; $(ii)$ Does OBA target users differently based on their profiles?; $(iii)$ Is OBA applied to sensitive topics?; $(iv)$ Is OBA more pronounced in certain geographic regions compared to others?; $(v)$ Do privacy configurations, such as Do-Not-Track, have any impact on OBA?.

Our methodology addresses all above challenges by 1) employing various \emph{filters} to distinguish interest-based targeting from other forms of advertising, 2) examining several alternative \emph{metrics} to quantify the extent of OBA advertising, 3) relying on multiple independent \emph{sources} to draw keywords and tags for the purpose of defining different interest types and searching for OBA around them, 4) allowing different geographical and privacy configurations. Our work combines all the above to present a much more complete methodology for OBA detection compared to very limited work existing in the area that has focused on particular special cases over the spectrum of alternatives that we consider (see Sec.~\ref{sec:rw} for related work).  

A second contribution of our work is the implementation and experimental application of our methodology. We have conducted extensive experiments for 72 interest-based personas (e.g., `motorcycles', `cooking', `movies', `poetry' or `dating') including typical privacy-sensitive profiles (e.g., `AIDS \& HIV', `left-wing politics' or `hinduism'), involving 3 tagging sources and 3 different filters. For each experiment we run 310 requests (on average) to 5 different context free ``test'' websites to gather more than 3.5M ads. Having conducted more than 2.9K experiments combining alternative interest definitions, geographical locations, privacy configurations, metrics, filters and sources of keywords to characterize OBA, we observe the following:

\vspace{2pt}

\noindent (1) OBA is a common practice, 88\% of the analyzed personas get targeted ads associated to all the keywords that define their behavioural trait. Moreover, half of the analyzed personas receive between 26-62\% of ads associated to OBA.

\noindent (2) The level of OBA attracted by different personas shows a strong correlation (0.4) with the value of those personas in the online advertising market (estimated by the CPC suggested bid for each persona).

\noindent (3) We provide strong evidences that show that the online advertising market targets behavioural traits associated to sensitive topics related to health, politics or sexual orientation. Such tracking is illegal in several countries \cite{eudirective}. Specifically, 10 to 40\% of the ads shown to  half of the 21 personas configured with a sensitive behavioural trait  correspond to OBA ads.

\noindent (4) We repeat our experiments in both US and Spain and do not observe any significant geographical bias in the utilization of OBA. Indeed, the median difference in the fraction of observed OBA ads by the considered personas in US and Spain is 2.5\%. 

\noindent (5) We repeat our experiments by having first set the Do-Not-Track (DNT) flag on our browser and do not observe any remarkable difference in the amount of OBA received with and without DNT enabled. This lead us to conclude that support for DNT has \emph{not yet} been implemented by most ad networks and sites.

%
%
%
%


Our intention with this work is to pave the way for developing a robust and scalable methodology and supporting toolsets for detecting interest-based targeting. By doing so we hope to improve the transparency around this important issue and protect the advertising ecosystem from the aforementioned Tragedy of the Commons. 

The rest of the paper is organized as follows: In Sec.~\ref{sec:behavioural} we describe what Online Behavioural Advertising is. Sec.~\ref{sec:methodology} describes the proposed methodology to unveil and measure the representativeness of OBA. Using this methodology we have implemented a real system that is described and evaluated in Sec.~\ref{sec:system}. Sec.~\ref{sec:results} presents the results of the conducted experiments. Finally, Sec.~\ref{sec:rw} analyzes the related work and Sec.~\ref{sec:conclusion} concludes the paper.  

\vspace{-6mm}

\section{Online Behavioural Advertising}
\label{sec:behavioural}

Online Behavioural \emph{targeted} Advertising (OBA) is the practice in online advertising wherein information about the interests of web users 
is incorporated in tailoring ads. This information 
is usually collected over time by aggregators or ad-networks while users browse the web. This information can 
include the publishers/ webpages a user browses as well as information on activity on each page (time spent, clicks, interactions, etc.). 
Based on the overall activity of the users, profiles can be built and these profiles can be used to increase the effectiveness of ads, 
leading to higher click-through rates and in turn, higher revenues for the publisher, the aggregator and eventually the advertiser 
by making a sale. We note that such targeting is referred to as network based targeting in the advertising literature.

 \begin{figure}[tb]
       \begin{center}
          \includegraphics[width=0.7\columnwidth]{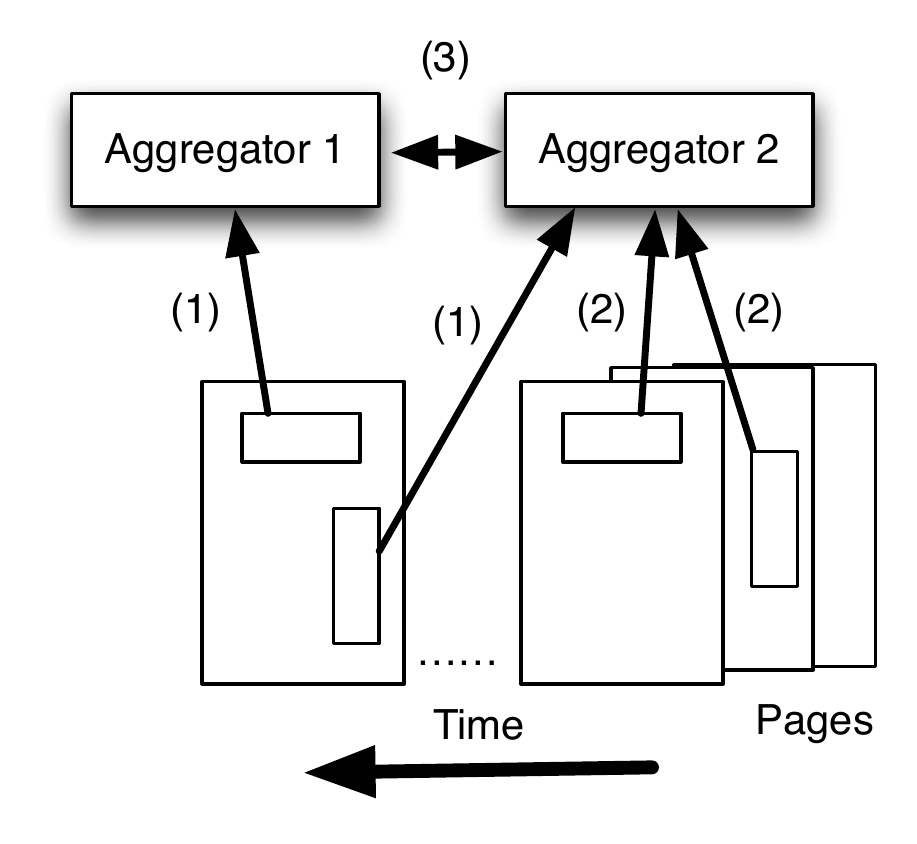}
      \end{center}
          \vspace{-8mm}
          \caption{\small High level description of how OBA can happen: User browses multiple webpages over time, each page 
has ads and aggregators present on them. When the user is on the current page (1), the aggregators present on that page can either 
be new or aggregators that were present on previous pages (2). Hence
      Aggregator 2 can leverage on past information to show a tailored ad, while Aggregator 1 can either show a 
run-of-network (RoN) ad or get information from Aggregator 2 (3) to show a tailored ad.}
          \label{fig:adnwk}
          \vspace{-3mm}
  \end{figure}

In Figure~\ref{fig:adnwk}, we provide a very high-level overview of how OBA can happen, and information gleaned by browsing can be used. 
Assume user Alice has no privacy protection mechanisms enabled in her browser. As she is visiting multiple 
publishers (e.g., websites), her activity is being tracked by multiple aggregators that are present on each publisher, using any of the available methods for tracking users~\cite{Gill2013, KrishnamurthyW09}. When Alice visits a publisher, aggregators (aggregator 2) present on that publisher could have already tracked her across the web and based on what information they have about her, they can target her accordingly. Another scenario can be when an aggregator (aggregator 1) is present on the current publisher where Alice is but was not present on previous publishers. In this case, the aggregator can either show a run-of-network ad (un-tailored) or obtain information about Alice from other aggregators and/or data sellers to show tailored ads. Indeed the full ecosystem consisting
of aggregators, data sellers, ad-optimizers, ad-agencies etc. is notoriously complex\footnote{\url{http://www.displayadtech.com/the_display_advertising_technology_landscape#/the-display-landscape}}~\cite{Shuai:AdvertisingSurvey}, however for the purposes of this work, we represent all entities handling data other than the user and the publishers, either collecting or selling data, as aggregators.

Other types of (less privacy intrusive) targeted advertising techniques include: $(i)$ \emph{Geographical Targeted Ads} are shown to a user based on its geographical location; $(ii)$ \emph{Demographic Targeted Ads} are shown to users based on their demographic profile (age, sex, etc) that is estimated by aggregators in different manners, for instance, through the user's browsing history \cite{barford2014adscape}; $(iii)$ \emph{Re-targeting Ads} present to the user recently visited websites, e.g., a user, that has checked a hotel in website A, receives an ad of that hotel when visiting website B few hours latter. Finally, a user can be exposed to \emph{Contextual Ads} when visiting a website. These ads are related to the theme of the visited website rather than the user's profile and thus we consider them as non-targeted ads.

%% file: methodology_v2.tex

 \vspace{-3mm}

\section{Methodology to measure OBA}
\label{sec:methodology}

In this section we describe our methodology to unveil the presence (or absence) of OBA advertising as well as to estimate its frequency and intensity compared to more traditional forms of online advertising. 

\vspace{-1mm}
\subsection{Rationale and Challenges}

Our goal is to uncover causal links between users exhibiting a certain behavioural trait and the display ads shown to them. Notice that we \emph{do not} claim or attempt to reverse engineer the complex series of online auctions taking place in real time. We merely try to detect whether there is any correlation between the advertisements displayed to a user and his past browsing behaviour. 

We create artificial \emph{personas}  that present 
a very narrow web browsing behaviour that corresponds to a very specific interest (or theme), e.g., `motor sports' or `cooking \& recipes'. 
We train each persona by visiting carefully selected websites 
that match its interest and by doing so invite data aggregators and trackers to classify our persona accordingly. 
We refer to the visited websites as \emph{training} webpages. For instance, the training set for the `motor sports' persona 
would be formed by specific motor sports webpages. Therefore, two first challenges for our methodology are which personas to examine and how to select training webpages for them that lead to a minimal \emph{profile contamination}~\cite{barford2014adscape}. By contamination, we are referring to the association of tags and labels not related to the main theme of the persona.  

Once the personas and the training webpages have been properly selected, we need to retrieve the ads that these personas obtain upon visiting carefully selected \emph{control} webpages that meet the following criteria: $(i)$ include a sufficient number of display ads, and $(ii)$ have a neutral or very well defined context that makes it easy to detect context based advertisements and filter them out to keep only those that could be due to OBA. We use weather related webpages for this purpose.  

The ads shown to a persona in the control pages lead to websites that we refer to as \emph{landing} webpages. Therefore if the theme of the landing webpages for a persona has a large overlap with the theme of its training websites  we can conclude that this persona frequently receives OBA ads. To automate and scale the estimation of the topical overlap between training and landing pages, we rely on online tagging services (e.g., Google AdWords, Cyren, etc) that categorize webpages based on keywords. We use them to tag each training and landing webpage associated to a persona and compute the existing overlapping. 
Note that we decided to  use several online tagging services or \emph{sources} to remove the dependency on a single advertising platform (a limitation of previous works like in~\cite{barford2014adscape, liu2013adreveal}).

As indicated before OBA can co-exist with several other types of advertisement on the same page, including: re-targeting ads, contextual ads, geographically targeted ads. Our goal is to define a flexible methodology able to detect and measure OBA in the presence of such ads. Thus, the fourth challenge that our methodology faces is to define filters for detecting and removing these other types of ads. 

The final challenge for our methodology is to define meaningful, simple, and easy to understand and measure metrics to quantify OBA using keywords from the training and landing pages of a persona.

\subsection{Details of the Methodology}
\label{subsec:meth_details}

\noindent\textbf{- Selection of Personas \& Training Pages:} 
The selection of personas with a very specific behavioural trait, and thus a reduced profile contamination, is a key aspect of our methodology. To achieve this in an automated and systematic manner we leverage the Google Ad Words' hierarchical category system, which includes more than 2000 categories that correspond to specific personas (i.e., behavioural traits) used by Google and its partners to offer OBA ads.  For each one of these personas, Ad Words provides a list of related websites (between 150 and 330 webs). We use these websites as training pages for the corresponding persona. 
Specifically, we consider the 240 personas of level 2 from the Google Ad Words' category system and apply the following three-steps filtering process to collect keywords for each persona while catering to avoid profile contamination:

\noindent \emph{- Step 1:}
For a given persona $p$, we collect the keywords assigned by Ad Words to every one of its related websites and keep in our dataset only those websites that have $p$'s category among their keywords. For instance, for $p$ = `motor sports', we only keep those related websites categorized by Ad Words with the keyword `motor sports'. After applying this step 202 personas remain in our dataset.

\noindent \emph{- Step 2:}
For each persona $p$, we visit each website selected during \emph{Step 1}, using a clean, in terms of configuration, browser, \ie, without cookies, previous web-browsing history, or ads preferences profile. Then, we check the categories from the Google Ad Words system added to the ads preference profile\footnote{Google's ads preferences profile represents the behavioural trait inferred by Google for a browser, based on the previously visited sites. It includes one or more categories from the Ad Words category system.} of our browser after visiting those websites. We only keep those related websites, which add to the ads preference profile exclusively $p$'s category or $p$'s category plus a second related one. For instance, for $p$ = `motor sports', we only keep a related website in our dataset if it includes to the ads preferences profile the category `motor sport' or the category `motor sports' plus a second one such as `cars' or `motorbikes'. After applying this step 104 personas remain in our dataset.

\noindent \emph{- Step 3:} Our final dataset consists of 51 personas that are left with at least 10 training pages each after steps 1 and 2. Note that by having at least 10 training pages we intend to capture enough diversity in the visited sites and expose our personas to a number of trackers that would approximate what a real user would find. Indeed, the considered personas are exposed to 15-40 trackers whereas the examination of the browsers of 5 volunteers revealed that they were exposed to 18-27 trackers\footnote{ We have used the tracker detection tool provided by the EDAA \cite{trackers} to obtain these results.}.
\vspace{1pt}

In addition to the above systematically collected personas, we have also selected manually 21 \emph{sensitive personas}  related to topics that, for instance,  privacy regulation in Europe do not allow to track or process (e.g., health, ethnicity, sexuality, religion or politics) \cite{eudirective}. Interestingly the categories of our \emph{sensitive personas} do not appear in the public Google Ad Words' Hierarchical Category System, however when querying for them in Ad Words we obtain a similar output as for any other persona. We apply the same steps as above with a single difference in \emph{Step 2}, where we keep only websites that do not add any category in the ads preference profile of our browser. By doing so, we ensure that our \emph{sensitive personas} are not being associated with any additional behavioural trait.

The final list of 51 regular and 21 sensitive personas can be checked in Figure~\ref{fig:bailp_and_TTK} and Figure~\ref{fig:bailp_and_TTK_SEN}, respectively. 

\vspace{2pt}

\noindent \textbf{- Selection of Control Pages:} As indicated in the methodology's rationale we need a set of pages that are popular, have ads shown on them and yet have low number of easily identifiable tags associated with them and thus do not contaminate the profile of our personas. We used five popular weather pages\footnote{\url{http://www.accuweather.com},\url{http://www.localconditions.com}, \url{http://www.wunderground.com}, \url{http://www.myforecast.com}, \url{http://www.weatherbase.com}} as control pages since they fulfil all previous requirements.
 

\noindent\textbf{- Visiting Training and Control Pages to obtain ads:}
Once we have selected the set of training and control pages for a persona, we visit them with the following strategy (see Sec.~\ref{sec:sys_implementation} for details). We start with a fresh install, and select  randomly a page from the pool of training+control pages to visit with the interval between different page visits drawn from an exponential distribution with mean 3 mins\footnote{This distribution is selected to emulate a human-being generated inter-arrival time between visits according to recent measurement studies \cite{Kumar:2010:COB:1772690.1772748}.}. By doing so, on the one hand, we regularly visit the training pages so that we allow trackers and aggregators present in those pages to classify our persona with a very specific interest according to our deliberately narrow browsing behaviour. On the other hand, the regular visits to control pages allow us to collect the ads shown to our persona to latter study whether they are driven by OBA. An alternative strategy would be to visit first the training pages several times to get our persona profiled by aggregators and visit only control pages. We avoided this strategy because visiting consecutively multiple weather sites fooled the data aggregators into believing that our browser was a ``weather'' persona.

\vspace{2pt}

\noindent\textbf{- Tagging Training and Landing Pages:}
In order to be able to detect systematically correlations between training and landing pages we need to first identify the keywords that best describe each webpage in our dataset. For this purpose, we use 3 different sources: Cyren\cite{cyren}, Google Ad Words\cite{google} and McAfee\cite{mcafee}. Each source has its own labeling system: Google Ad Words labels web-pages using a hierarchical category system with up to 10 levels and tag categories with 1 to 8 keywords. Cyren and McAfee  provide a flat tagging system consisting of 60-100 categories  and label web-pages with at most 3 keywords. Note that by utilising multiple sources we try to increase the robustness of our methodology and limit as much as possible its dependency to the idiosyncrasies of particular labeling systems.
Finally, it is worth noting that the coverage of the considered tagging services is very high for our set of training and landing pages. In particular Google, McAfee and Cyren were able to tag 100\%, 99.0\% and 95.5\% of the training pages and 100\%, 97.2\% and 93.3\% of the landing pages, respectively.

\vspace{2pt}

\noindent\textbf{- Training Set Keywords:}
To achieve the aforementioned robustness against the particularities of individual classification systems, we filter the keywords assigned to a page by keeping only those that are assigned to the page by more than one of our 3 sources. The idea is to quantify OBA based on keywords that several of our sources agree upon for a specific page relevant to the trained persona.
Assume we have a training webpage $W$ tagged with the set of keywords $K_1$ to $K_3$ for each one of the 3 sources above. Our goal is to select a keyword $k$ within $K_i$ ($i \in [1,3]$) only if it accurately defines $W$ for our purpose. To do this, we leverage  the Leacock-Chodorow similarity \cite{leacock1998using} ($S(k,l)$) to capture how similar two word senses (keywords) $k \in K_i$ and $l \in K_j$ ($j \in [1,3]~\&~j \neq i$) are. Note that two keywords are considered similar if their Leacock-Chodorow similarity is higher than a given configurable threshold, $T$, that ranges between 0 (any two keywords would be considered similar) and 3.62 (only two identical keywords -exact match- would be considered similar). We compute the similarity of $k$ belonging to a given source with all the training keywords belonging to other sources and consider $k$ an accurate keyword only if it presents a $S(k,l) > T$ with keywords of at least $N$ other sources. Note that $N$ is also a configurable parameter that allows us to define a more or less strict condition to consider a given training keyword in a given source.

\vspace{2pt}

\noindent\textbf{- Filtering different types of ads:} 
To complete our methodology we describe next the filters used in order to identify and progressively remove landing pages associated with non-OBA ads:

\noindent \emph{- Retargeting Ads Filter ($F_{r}$)}: In our experiment a retargeting ad in a control page should point to either a training or a control page previously visited by the persona. Since, we keep record of the previous webpages visited by a persona, identifying and removing retargeting ads from our landing set is trivial.

\noindent \emph{- Static and Contextual Ads Filter ($F_{s\&c}$)}: We have created a profile that after visiting each webpage removes all cookies and potential tracking information such that each visit to a website emulates the visit of a user with empty past browsing history. We refer to this persona as \emph{clean profile}. By definition, when visiting a control webpage the clean profile cannot receive any type of targeted behavioural ad and thus all ads shown to this profile correspond to either static ads (ads pushed by an advertiser into the website) or contextual ads (ads related to the theme of the webpage). Hence, to eliminate a majority of the landing pages derived from static and contextual ads for a persona, we remove all the common landing pages between this persona and the clean profile.

\noindent \emph{- Demographic and Geographical Targeted Ads Filter ($F_{d\&g}$)}: We launch the experiments for all our personas from the same /24 IP prefix and therefore it is likely that several of them receive the same ad when geographical targeting is used. Moreover, we have computed the  Leacock-Chodorow similarity between the categories of each pair of personas in our dataset to determine how close their interests are.
To filter demographic and geographical ads we proceed as follows: for a persona $p$ that has received an ad $A$, we select the set of other personas receiving this same ad ($O(A) = [p_{1,A}, p_{2,A},...]$) and compute the Leacock-Chodorow similarity between  $p$ and $p_{i,A} \in O(A)$. If the similarity between $p$ and at least one of these personas is lower than a given threshold $T'$, we consider that the ad has been shown to personas with a significantly different behavioural trait and thus it cannot be the result of OBA. Instead, it is likely due to geographical or demographic targeting practices.

\vspace{2pt}

\noindent\textbf{- Measuring the presence and representativeness of OBA:}
\label{sec:metrics} We measure the volume of OBA for a given persona $p$ by computing the overlapping between the keywords of the training and landing pages for $p$. Note that we consider that a training keyword and a landing keyword overlap if they are an exact match. In particular, we use two complementary metrics that measure different aspects of the overlapping between the keywords of training and landing pages. However, let us first introduce some definitions used in our metrics:  \textit{(i)} We define the set of unique keywords associated with the training pages for a persona \textit{p} on source \textit{s} as \textit{K$_{T_{ps}}$}; \textit{(ii)} We define the set of unique keywords associated with the landing pages of ads shown to a persona \textit{p} on source \textit{s} on control pages as \textit{K$_{L_{ps}}$}; \textit{(iii)} Finally we define the set of unique keywords associated to a single webpage \textit{W} on source \textit{s} as \textit{K$_{W_{s}}$}. Note that the set of keywords associated to a web-page remains constant for a given source regardless the persona. Using these definitions we define our metrics as follows:

\vspace{0.1cm}
\noindent \emph{Targeted Training Keywords (TTK):} This metric computes the fraction of keywords from the training pages that have been targeted and thus appear in the set of landing pages for a persona $p$ and a source $s$. It is formally expressed as follows:

\vspace{-0.1cm}
\begin{equation}
TTK(p,s) = \dfrac{|K_{T_{ps}} \wedge K_{L_{ps}}|}{|K_{T_{ps}}|} \in [0,1]
\end{equation}

In essence, TTK measures whether $p$ is exposed to OBA or not. In particular, a high value of TTK indicates that most of the keywords defining the behavioural trait of $p$ (i.e., training keywords) have been targeted during the experiment.


\vspace{0.2cm}

\noindent \emph{Behavioural Advertising in Landing Pages (BAiLP):} This metric captures the fraction of ads whose landing pages are tagged with at least one keyword from the set of training pages for a persona $p$ and a source $s$. In other words, it represents the fraction of received ads by $p$ that are likely associated to OBA. BAiLP is formally expressed as follows:

\begin{equation}
\begin{gathered}
BAiLP(p,s) = \dfrac{\sum_{i=1}^{L_{ps}}f(K_{W^{i}_{s}})\cdot ntimes}{L_{ps}}  \in [0,1] \\
where \: f(K_{W^{i}_{s}}) = \left\{\begin{matrix}
1 & if & (K_{T_{ps}} \wedge K_{W^{i}_{s}}) \geq 1\\ 
0 & if & (K_{T_{ps}} \wedge K_{W^{i}_{s}}) = 0
\end{matrix}\right.
  \end{gathered}
\end{equation} 
 
Note that \textit{ntimes} represents the number of times an ad has been shown to \textit{p} and \textit{L$_{ps}$} is defined as the set of landing pages for a persona \textit{p} and source \textit{s}.

\vspace{0.2cm}

In summary, TTK measures if OBA is happening and how intensely (what percentage of the training keywords are targeted) whereas BAiLP captures what percentage of the overall advertising a persona receives is due to OBA (under different filters).

%% file: system.tex
\section{Automated System to measure OBA} 
\label{sec:system}

In this section we describe the development and evaluation of a system that we developed for implementing our previously described methodology for measuring OBA.

\subsection{System implementation and setup} 
\label{sec:sys_implementation}

A primary design objective of our measurement system was to be fully automated, without a need for man-in-the-loop, in any of its steps. The reason for this is that we wanted to be able to check arbitrary numbers of personas and websites, instead of just a handful. Towards this end, we used a lightweight, headless browser PhantomJS ver. 1.9 (http://phantomjs.org/) as our base as we can automate collection and handling of content as well as configure different user-agents. We wrote a wrapper around PhantomJS that we call PhantomCurl that handles the logic related to collection and pre-processing of data. Our control server was setup in Madrid, Spain. The experiments were run from Spain and United States. In the case of US we used a transparent proxy with sufficient bandwidth capacity to forward all our requests. We used a user-agent\footnote{We have repeated some experiments using different user-agents without noticing major differences in the obtained results.} corresponding to Chrome ver. 26, Windows 7. Our default setup has no privacy protections enabled for personas, but for the clean profile, we enable privacy protection and delete cookies after visiting each web-site. A second configuration set-up enables the Do Not Track\footnote{Do Not Track is a technology and policy proposal that enables users to opt out of tracking by websites they do not visit (e.g., analytics services, ad networks, etc).} \cite{dnt} for all our personas. For each persona  configuration (no-DNT and DNT) and location (ES and US) we run, in parallel, our system 4 times in slots of 8-hours in a window of 3 days so that all personas are exposed to the same status of the advertising market. These time slots generate 310 visits per persona (on average) to the control pages that based on the the results in \cite{barford2014adscape} suffices to obtain the majority of distinct ads received by a persona in the considered period of time.
To process the data associated to each persona, configuration and geographical location we use 3 sources to tag the training and landing pages (Google, McAfee and Cyren), 3 different combinations of filters ($F_{r}$; $F_{r}$ and $F_{s\&c}$; $F_{r}$, $F_{s\&c}$ and $F_{d\&g}$) and 2 metrics (TTK and BAiLP). Furthermore, we use different values of T and N (for the selection of training keywords) and  T' (for filtering out demographic and geographic targeted ads). Overall our analysis covers more than 2.9K points in the spectrum of interest definitions, metrics, sources, filters, geographical locations, privacy configurations, etc.

Before discussing the obtained results (Sec.~\ref{sec:results}), in the next subsection we evaluate the performance of our system to identify OBA ads using standard metrics such as accuracy, false positive ratio, false negative ratio, etc. It is worth mentioning that, to the best of our knowledge, previous measurement works in the detection of OBA \cite{barford2014adscape, liu2013adreveal} do not perform a similar evaluation of their proposed methodologies.


\subsection{System Performance to identify OBA ads}
\label{sec:sys_efficiency}

To validate our system we need to generate a ground truth dataset to compare against. We used humans for a subjective validation of correlation between training and landing pages as done also by previous works \cite{Carrascosa2013COSN,lecuyer2014xray,TTclassif,UNED}.  To that end, two independent panelists subjectively classified each one of the landing pages associated to few\footnote{Note that the manual classification process required  our panelists to carefully evaluate around 300-400 landing pages per persona. Then, it was infeasible to perform it for every persona.} randomly selected personas as OBA or non-OBA. Note that the classification of these two  panelists was different in 6-12\% of the cases for the different personas. For these few cases a third person performed the subjective classification to break the tie.

For each ad, we compare the classification done by our tool as OBA vs. non-OBA with the ground truth and compute widely adopted metrics used to evaluate the performance of detection systems: Recall (or Hit Ratio), Accuracy, False Positive Rate (FPR) and False Negative Rate (FNR). Table \ref{tab:performance} shows the max and min value of these metrics across the analyzed personas for our three sources (McAfee, Google and Cyren). 
We observe that, in general, our system is able to reliably identify OBA ads for all sources. Indeed, it shows Accuracy and Recall values over 94\% as well as a FNR smaller than 4.5\%. Finally, the FPR stay lower than 10\% for all the analyzed personas excepting for the `Yard \& Patio' persona where the FPR increases up to 25\%.  


\begin{table}[t]
\centering
\scriptsize
\begin{tabular}{l|c|c|c|c}
 & \textbf{Recall} & \textbf{Accuracy} & \textbf{FPR}  &\textbf{FNR} \\ \hline
\textbf{McAfee} &99.1/95.6\% & 99.0/94.2\% & 25.5/3.8\% & 4.4/0.1\%  \\ 
\textbf{Google} &99.1/95.7\% & 99.0/94.3\% & 25.7/3.8\% & 4.3/0.1\%  \\
\textbf{Cyren} & 98.7/95.5\% & 98.6/94.1\%  & 25.4/4,27\% & 4.5/1,3\%\\ 
\end{tabular}
\caption{\small Max and Min values of Recall, Accuracy, FPR and FNR of our automated methodology to identify OBA ads for our three sources (McAfee, Google and Cyren) for the analyzed personas. Max and Min values correspond to `Bycicles Accesories' and `Yard \& Patio' personas, respectively.}
\label{tab:performance}
\end{table}



%% file: results3.tex
\section{Measuring OBA}
\label{sec:results}

In this section we present the results obtained with our measurement system for the purpose of answering the following essential questions regarding OBA $(i)$ How frequently is OBA used in online advertising?; $(ii)$ Does OBA target users differently based on their profiles?; $(iii)$ Is OBA applied to sensitive topics?; $(iv)$ Is OBA more pronounced in certain geographic regions compared with others?; $(v)$ Does Do-Not-Track have any impact on OBA?

We will start by analysing a concrete example and try to help the reader follow along the different steps of our methodology. After that we will present holistic results from a large set of experiemtns.


\subsection{Specific case: Swimming Pools \& Spas and Google}

Let us consider a persona, `Swimming Pools \& Spas', and a source, `Google', to present the results obtained in each step of our methodology for this specific case.
Table \ref{tab:training} shows the set of training webpages for the `Swimming Pools \& Spas' persona. One can observe by the name of the webpages their direct relation to the `Swimming Pools \& Spas' persona.

\begin{table}[t]
\scriptsize
\centering
\begin{tabular}{|ll|}
\hline
\multicolumn{2}{|c|}{\textbf{Training Pages}} \\\hline\hline
http://poolpricer.com & http://levelgroundpool.com \\ 
http://whirlpool-zu-hause.de & http://poolforum.se\\
http://eauplaisir.com & http://photopiscine.net \\ 
http://a-pool.czm & http://allas.fi \\ 
http://seaglasspools.com & http://piscineinfoservice.com\\\hline
\end{tabular}
\caption{\small Sample of the training webpages for `Swimming Pools \& Spas' persona}
\label{tab:training}
\end{table}

We train this persona as described in Section \ref{subsec:meth_details}. Then, in the post-processing phase we tag the training and landing webpages using our 3 sources. To describe the process in this subsection we refer to the results obtained for `Google'.
Table \ref{tab:training2} shows the 6 keywords that Google assigns to the training websites in the first column. However, this initial set of keywords may present some contamination including keywords unrelated to the `Swimming Pools \& Spas' persona. Hence, we compute the semantic similarity between these keywords and the keywords assigned by other sources to the training webpages with N = 2 and T = 2.5. This technique eliminates 2 keywords and leaves a final set of 4 training keywords shown in the second column of Table \ref{tab:training2}. 

\begin{table}[t]
\scriptsize
\centering
\begin{tabular}{l|l}
\multicolumn{1}{c|}{\textbf{Training Keywords}} & \multicolumn{1}{c}{\textbf{Filtered Training}} \\
& \multicolumn{1}{c}{\textbf{Keywords}}\\ \hline
Gems \& Jewellery & \multicolumn{1}{c}{---} \\ 
Gyms \& Health Clubs & \multicolumn{1}{c}{---} \\ 
Outdoor Toys \& Play Equipment & Outdoor Toys \& Play Equipment \\ 
Security Products \& Services & Security Products \& Services \\ 
Surf \& Swim & Surf \& Swim \\ 
Swimming Pools \& Spas & Swimming Pools \& Spas \\  
\end{tabular}
\caption{\small Keywords associated to the training websites for `Swimming Pools \& Spas' using Google as source before and after applying the semantic overlapping filtering with N = 2 and T = 2.5}
\label{tab:training2}
\end{table}

\begin{table}[t]
\scriptsize
\centering
\begin{tabular}{l|l|l}
& \textbf{TTK} & \textbf{BAiLP} \\ \hline
\textbf{$F_{r}$} & 1 & 0.17\\
\textbf{$F_{s\&c}$} & 1 & 0.75\\
\textbf{$F_{d\&g}$} & 1 & 0.97\\
\end{tabular}
\caption{\small TTK and BAiLP values after applying each filter for the `Swimming Pools \& Spas' persona and `Google' source.}
\label{tab:TTK_bailp_google_example}
\end{table}

\begin{table}[t]
\scriptsize
\centering
\begin{tabular}{l|c}
\multicolumn{1}{c|}{\textbf{Landing Webpages}} & \multicolumn{1}{c}{\textbf{Num. ads}} \\ \hline
www.abrisud.co.uk & 1195 \\ 
www.endlesspools.com & 106 \\ 
www.samsclub.com & 16 \\ 
www.paradisepoolsms.com & 8 \\ 
www.habitissimo.es & 8 \\ 
www.abrisud.es & 6 \\ 
ww.atrium-kobylisy.cz & 6 \\ 
www.piscines-caron.com & 5 \\ 
athomerecreation.net & 4 \\ 
www.saunahouse.cz & 4 \\ 
\end{tabular}
\caption{\small Top 10 list of landing webpages and the number of times their associated ads were shown to our `Swimming Pools \& Spas' persona.}
\label{tab:top10_babies}
\end{table}

\begin{figure*}[t]
\centering
\subfigure[TTK]{\includegraphics[width=\columnwidth]{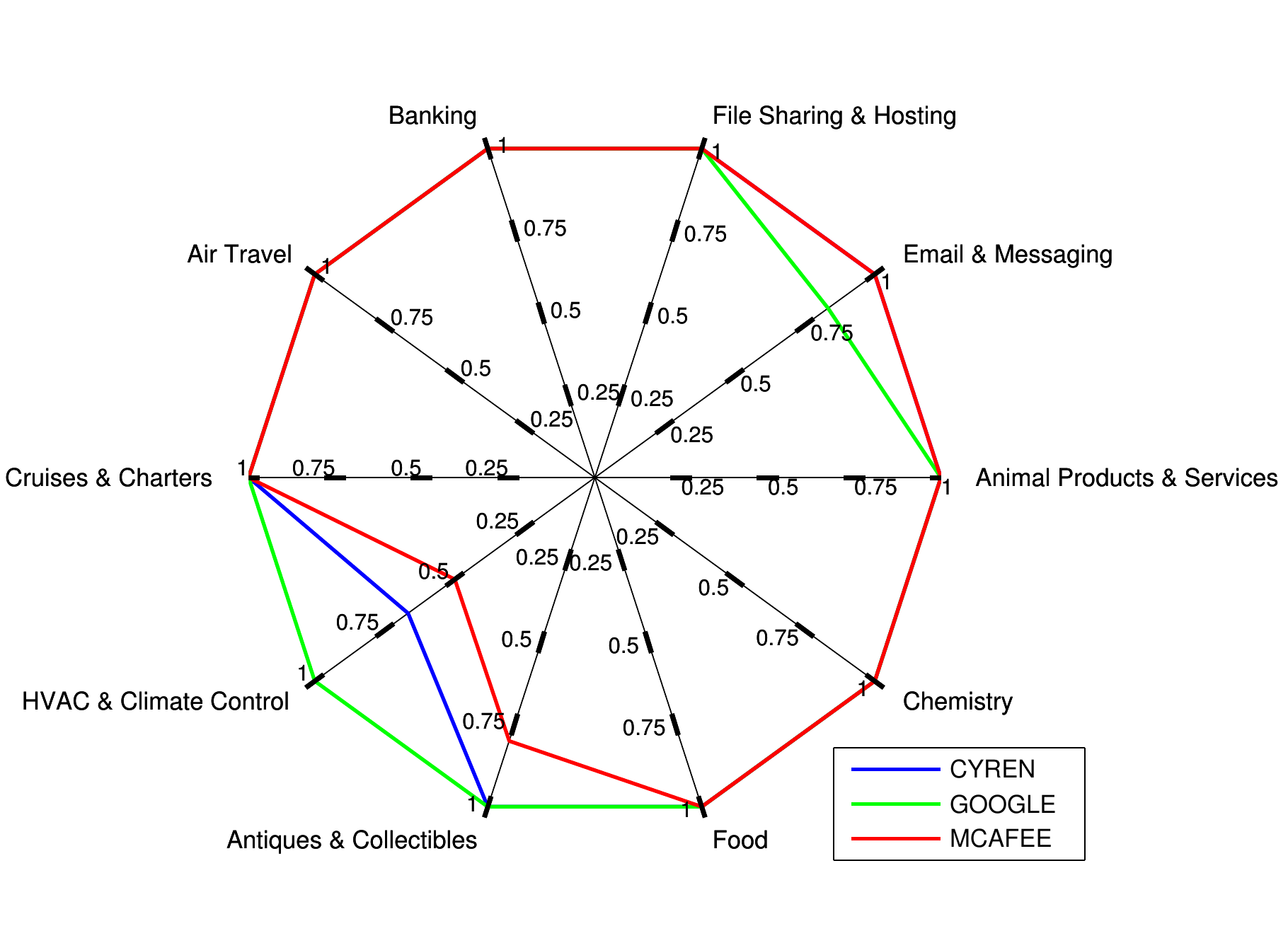} \label{subfig:boxplot_TTK}}\hfill
\subfigure[BAiLP]{\includegraphics[width=\columnwidth]{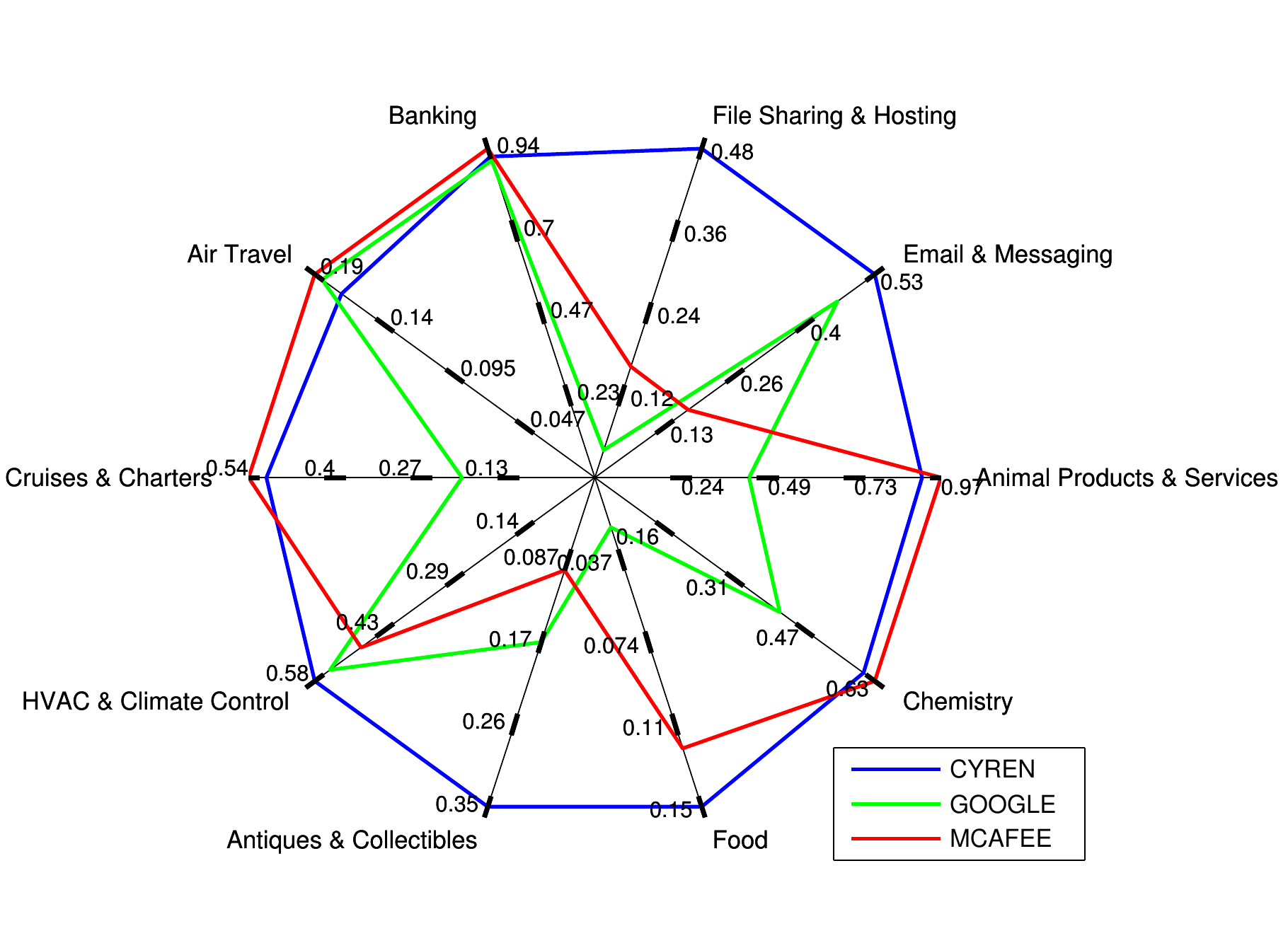} \label{subfig:boxplot_BAiLP}}
\caption{\small TTK and BAiLP for 10 personas and all sources for N = 2, T = T' = 2.5 and  all filters activated ($F_r$, $F_{s\&c}$,  $F_{d\&g}$)}
\label{fig:spider_charts}
\end{figure*}

\begin{figure*}[t]
\centering
\includegraphics[width=1.3\columnwidth]{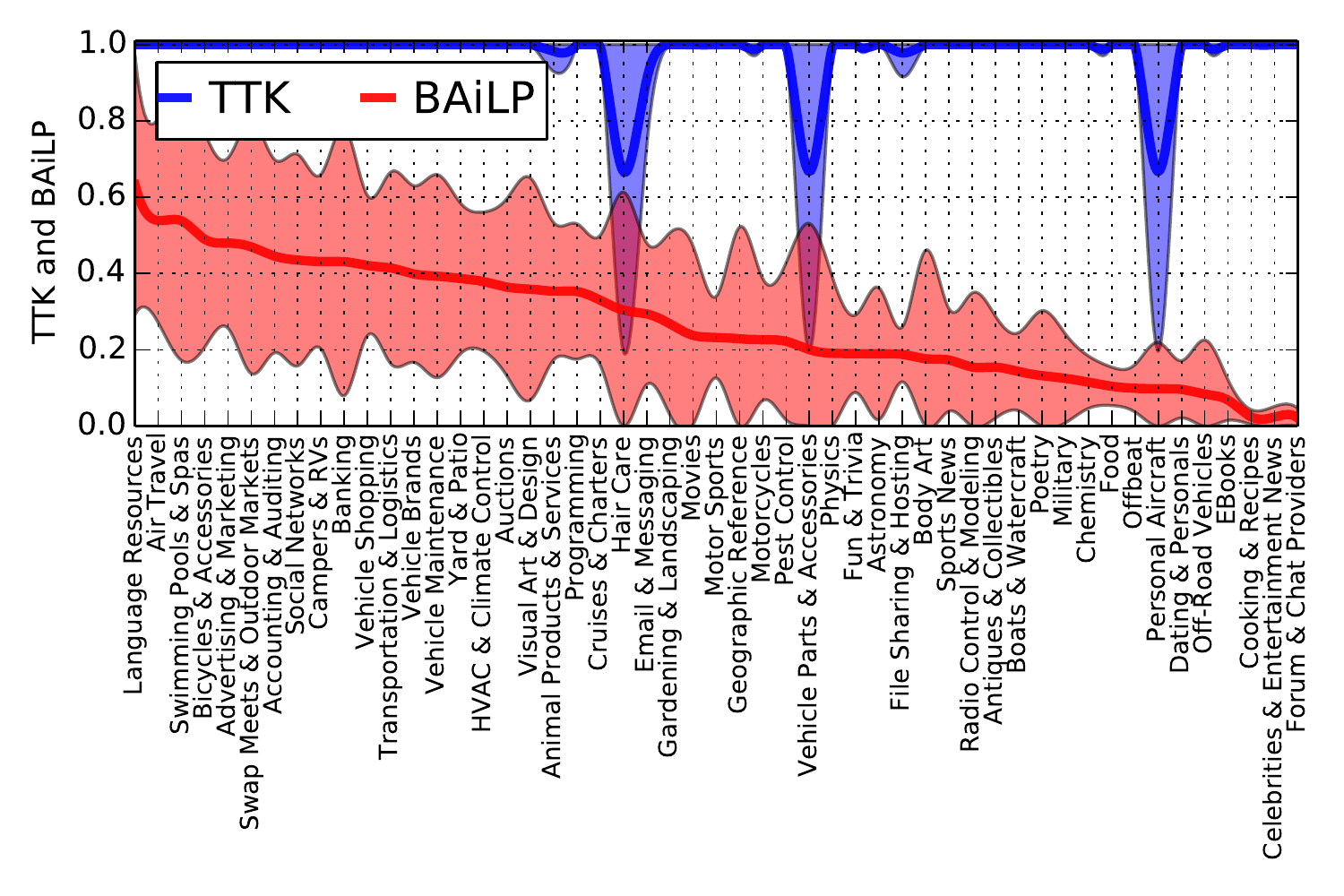} 
\vspace{-0.5cm}
\caption{\small Average and standard deviation of TTK and BAiLP for each regular persona in our dataset sorted from higher to lower average BAiLP.}
\label{fig:bailp_and_TTK}
\end{figure*}

\begin{figure*}[t]
\centering
\includegraphics[width=1.3\columnwidth]{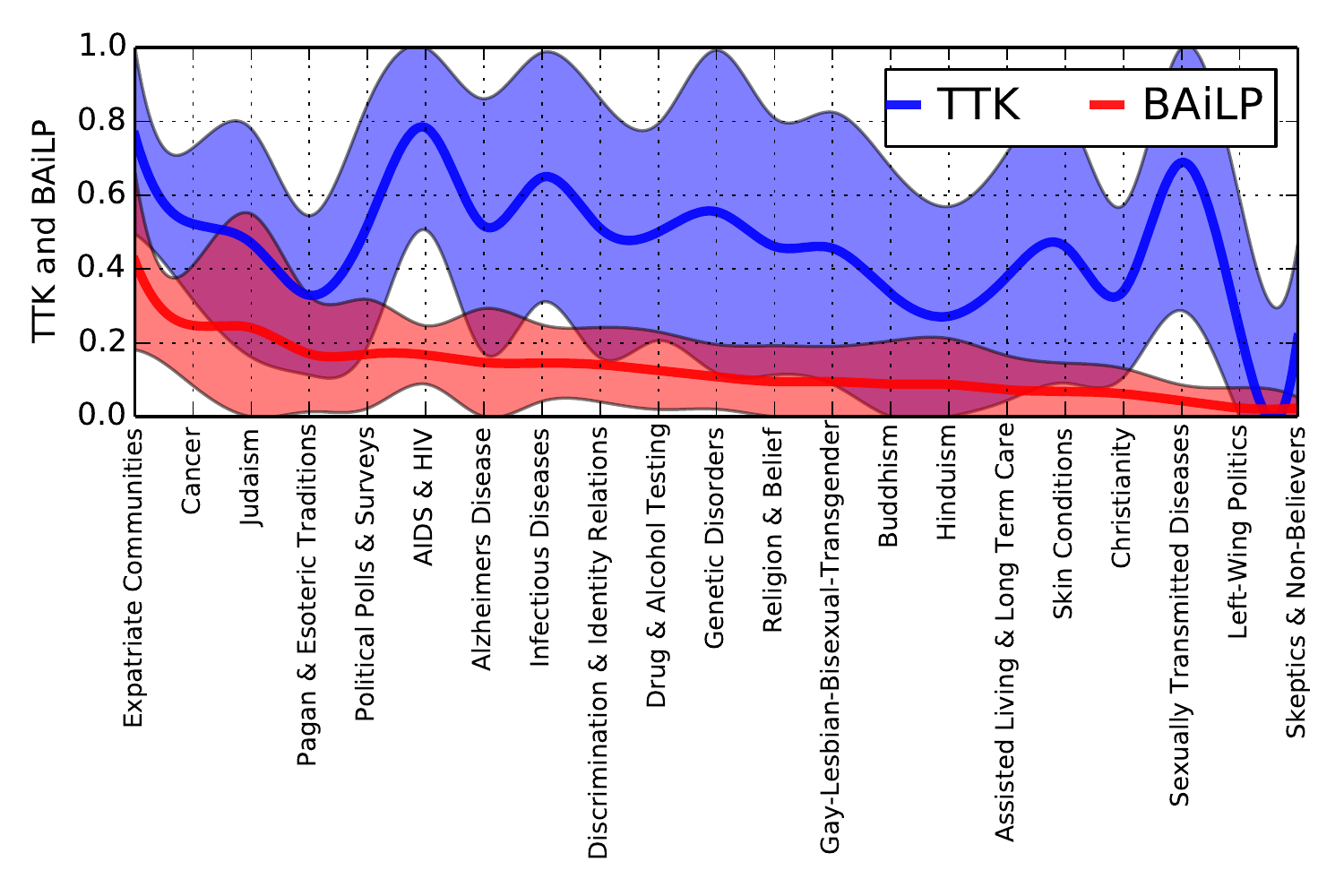} 
\vspace{-0.5cm}
\caption{\small Average and standard deviation of TTK and BAiLP for each sensitive persona in our dataset sorted from higher to lower average BAiLP.}
\label{fig:bailp_and_TTK_SEN}
\end{figure*}

Let us now focus on the landing webpages. Our experiments provide a total of 381 unique landing webpages after filter $F_{r}$. Then, we pass each of these webpages for  $F_{s\&c}$ and $F_{d\&g}$ filters sequentially. Each filter eliminates 226 and 128 of the initial landing pages, respectively. This indicates that contextual ads (eliminated by $F_{s\&c}$) are the more frequent type of  ads. After applying each filter we compute the value of the two defined metrics (TTK and BAiLP) using the resultant set of landing pages and its associated keywords and show them in Table~\ref{tab:TTK_bailp_google_example}. 
The results suggest a high presence of OBA ads. Indeed, the obtained TTK values indicate that 100\% of the training keywords are targeted and thus they appear among the landing keywords. Moreover, the BAiLP shows that, depending on the specific applied filter, between 17 and 97\% of received ads by our `Swimming Pools \& Spas' persona are associated to landing pages tagged with  keyword from the training set and thus are likely to be associated to OBA advertising. Note that in the extreme case where no filters are applied, BAiLP represents the percentage of all ads shown that are suspected to be targeted (17\% for `Swimming Pools \& Spas' persona). In the other extreme, when all filters are applied, BAiLP shows the same percentage after having removed all advertisements that can be attributed to one of the known categories described in Sec.~\ref{sec:behavioural} (97\% for `Swimming Pools \& Spas' persona).



Finally, Table \ref{tab:top10_babies} shows the Top 10 landing pages associated to a larger number of ads shown during our experiment. We observe that the three most frequent landing pages, that amount to most of the ads shown to our persona, are related to Swimming Pools, pointing clearly to OBA.

\subsection{How frequent is OBA?}
\label{subsec:freq}

Let us start analyzing the results obtained with our methodology for each independent source. For this purpose, Figure~\ref{fig:spider_charts} presents  the values of TTK and BAiLP for 10 selected personas in a radar chart. In particular, these results correspond to experiments run from Spain, with DNT disabled and  all filters ($F_{r}$, $F_{s\&c}$, $F_{d\&g}$)  activated.
First, TTK shows its maximum value (i.e., 1) in 9 of the studied personas for Google and Cyren and in 8 personas for McAfee. Moreover, TTK is not lower than 0.5 in any case. This result indicates that regardless of the source used to tag websites, typically all the training keywords of a persona are targeted and then appear in its set of landing keywords. Second, we observe a much higher heterogeneity  for BAiLP across the different sources. In particular, Cyren seems to consistently offer a high value of BAiLP in comparison with the other sources whereas McAfee offers the highest (lowest) BAiLP for 5 (3) of the considered personas and shows a remarkable agreement (BAiLP difference $<$ 0.05) with Cyren in half of the considered personas. Google is the most restrictive source offering the lowest BAiLP for 6 personas. In addition it only shows close agreement with Cyren and McAfee for two personas (`Air Travel' and `Banking').   These results are due to the higher granularity offered by Google compared to Cyren and McAfee that makes more difficult finding matches between training a landing keywords for that source. If we now compare the BAiLP across the selected personas, we observe that for 27 of the 30 considered cases BAiLP ranges between 0.10 and 0.94 regardless of the source. This indicates that 10-94\% of the received ads by these personas are associated to landing pages tagged with training keywords and then, they are likely to be the result of OBA.


These preliminary results suggest an important presence of OBA in  online advertising. In order to confirm this observation and  understand how representative OBA is, we have computed the values of our two metrics, TTK and BAiLP, for every combination of  persona, source, set of active filters and setting N = 2, T = T' = 2.5 in our dataset. Again these experiments are run from Spain and  with DNT disabled. In total 4 runs of 459 independent experiments were conducted. Figure~\ref{fig:bailp_and_TTK} shows the average and standard deviation values of TTK and BAiLP for the 51 considered personas, sorted from higher to lower average BAiLP value. Our results confirm a high presence of OBA ads. The obtained average TTK values indicate that for 88\% of our personas all training keywords are targeted and appear among the landing keywords (for the other 12\% at least 66\% of training keywords match their correspondent landing keywords). This high overlapping  shows unequivocally the existence of OBA. However to more accurately quantify its representativeness we rely on our BAiLP metric, which demonstrates that half of our personas are exposed (on average) to 26-63\% of ads linked to landing pages tagged with keywords from the persona training set. Since the overlap is consistently high, independently of the source, filters used, \etc, we conclude that these ads are likely the result of OBA.

\subsection{Are some personas more targeted than others?}

Figure~\ref{fig:bailp_and_TTK} shows a \emph{clear variability in the representativeness of OBA for different personas}. Indeed, the distribution of the average BAiLP values across our personas presents a median value equal to 0.23 with an interquartile range of 0.25 and a max/min value of 0.63/0.02. This observation invites the following question: \emph{``Why are some personas targeted more intensely than others?''}. Our hypothesis is that the level of OBA received by a persona depends on its economic value for the online advertising market. To validate this hypothesis we leverage the AdWords keyword planner tool\footnote{\url{https://adwords.google.com/KeywordPlanner}}, which enables us to obtain the suggested Cost per Click (CPC) bids for each of our personas.\footnote{Specifically, we use the keyword defining the interest of each persona to obtain its suggested CPC bid.} The bid value is a good indication of the relative economic value of each persona. 
Then, we compute the spearman and pearson correlation between the BAiLP and the suggested CPC for each persona in our dataset. Note that to properly compute the correlation, we eliminate  outlier samples based on the suggested CPC.\footnote{We use a standard outlier detection mechanism, which considers a sample as an outlier if it is higher (smaller) than Q3+1.5*IQR (Q1-1.5*IQR) being Q1, Q3 and IQR the first quartile, the third quartile and the interquartile range, respectively.}
The obtained spearman and pearson correlations are 0.44 and 0.40 (with p-values of 0.004 and 0.007), respectively. These results validate our hypothesis since \emph{we can observe a marked correlation between the level of received OBA (BAiLP) and the value of the persona for the online advertising market (suggested CPC bid)}.

\subsection{Is OBA applied to sensitive topics?}

The sensitive personas in our dataset present behavioural traits associated to sensitive topics including health, religion, and politics. Tracking these topics is illegal (at least) in Europe. To check if this is being respected by the online advertising market, we repeat the experiment described in the previous subsection for all our 21 sensitive personas, setting the geographical location in Spain. In this case, we run 4 repetitions of 189 independent experiments.

Figure~\ref{fig:bailp_and_TTK_SEN} shows the average and standard deviation values of TTK and BAiLP for each sensitive persona, sorted again from higher to lower average BAiLP value.
One would expect to find values of TTK and BAiLP close to zero indicating that sensitive personas are not subjected to OBA. Instead, our results reveal that despite the lower values compared to the personas of Figure \ref{fig:bailp_and_TTK}, the median value of average TKK is 0.47 indicating that for half of the sensitive personas at least 47\% of the  keywords defining their behavioural trait remain targeted. Moreover, BAiLP results show that 10-40\% of the ads received by half of our sensitive personas are associated to OBA. In summary, \emph{we have provided solid evidence that sensitive topics are tracked and used for online behavioural targeting despite the existence of regulation against such practices}.


\subsection{Geographical bias of OBA}

In order to search for possible geographical bias of OBA, we have run the 459 independent experiments described in Subsection \ref{subsec:freq} using a transparent proxy configured in US so that visited websites see our measurement traffic coming from a US IP address. For each persona, we have computed the average BAiLP across all combinations of sources and filters for the experiments run in Spain vs. US and calculated the BAiLP difference. Figure~\ref{fig:geo_and_dnt} shows the distribution of average BAiLP difference for the 51 considered personas in the form of a boxplot. Note that a positive (negative) difference indicates a major presence of OBA ads in Spain (US). The BAiLP differences are restricted to less than 10 percentage points across all cases with an insignificant bias (median of BAiLP difference = 2.5\%) towards a major presence of OBA ads in Spain than in US. Hence, we conclude that \emph{there is not a remarkable geographical bias in the application of OBA}. Note that we have repeated the experiment with our other metric, TTK, obtaining similar conclusions.

\subsection{Impact of Do-Not-Track in OBA}

\begin{figure}[t]
\centering
\includegraphics[width=0.75\columnwidth]{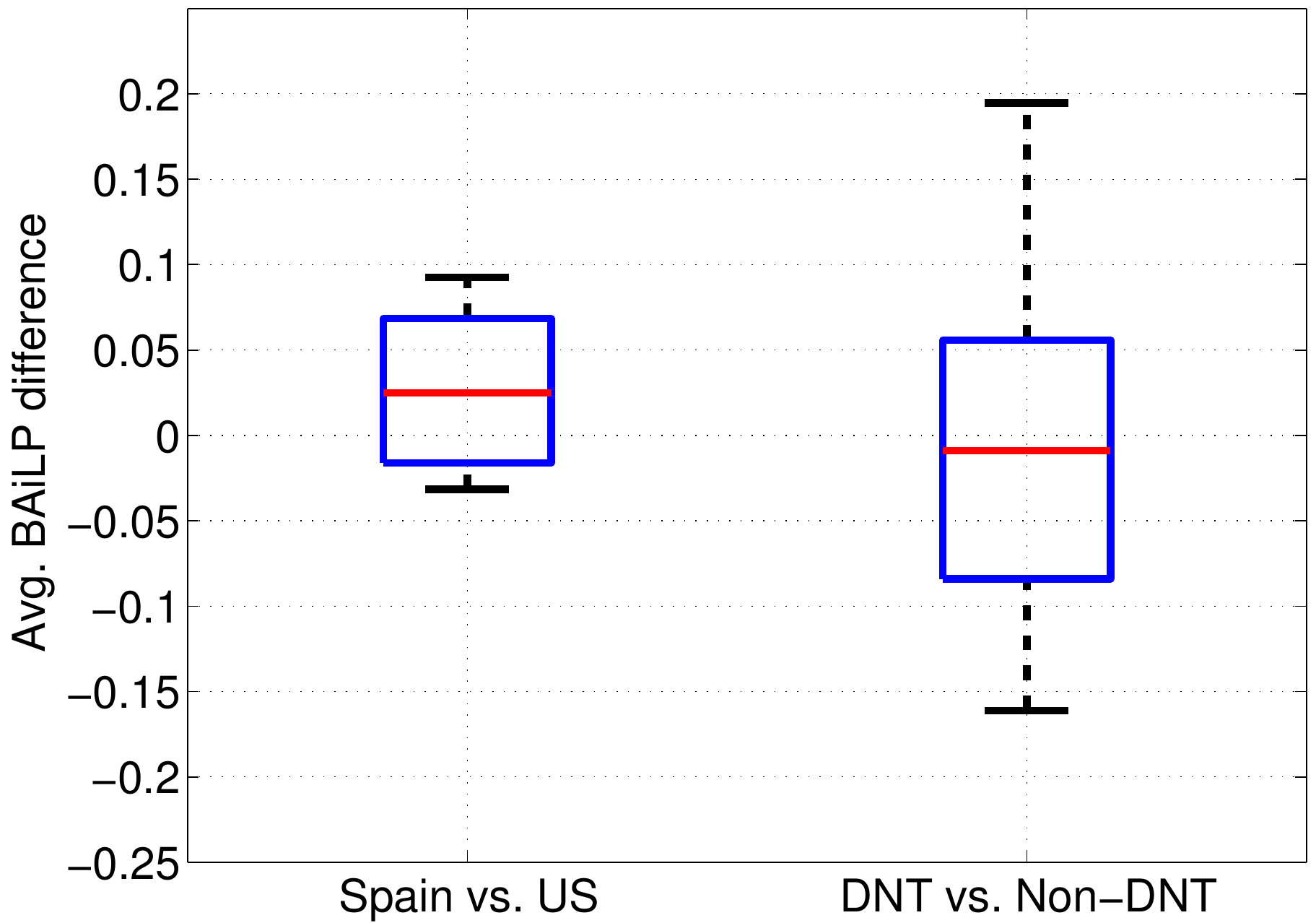} 
\caption{\small Distribution of the average BAiLP difference for the 51 regular personas in our dataset for the cases: Spain vs. US (left) and DNT vs. non-DNT (right)}
\label{fig:geo_and_dnt}
\vspace{-0.3cm}
\end{figure}

Following the same procedure as in the previous subsection, we have computed the average BAiLP difference when DNT is activated from when DNT is not for each one of the 51 regular personas in our dataset, fixing in both cases the geographical location to Spain. Figure~\ref{fig:geo_and_dnt} depicts the distribution of the average BAiLP difference where, a positive (negative) difference indicates a major presence of OBA ads with the DNT activated (deactivated). The median of the distribution is $\sim$0, indicating that half of the personas attract more OBA ads either with DNT activated or not. Moreover, the IQR reveals that half of the personas present a relatively small BAiLP difference ($\le$ 8 percentage points).
Therefore, \emph{the results provide strong evidences that DNT is barely enforced in Internet and thus its impact in OBA is negligible}. Again, we have repeated this experiment with TTK obtaining similar conclusions.

%% file: related_work.tex
\section{Related Work}
\label{sec:rw}

Our work is related to recent literature in the areas of measurement driven studies on 
targeting/personalization in online services such as search \cite{hannak-2013-filterbubbles, Majumder:2013:KYP:2488388.2488464} and e-commerce \cite{Mikians12:hotnets,Mikians13:conext, hannak-2014-ecommerce}. 
More specifically, in the context of advertising the seminal work by Guha et al.~\cite{guha2010challenges} presents the challenges of measuring targeted advertising, including high levels of noise due to ad-churn, network effects like load-balancing, and timing effects. 
Our methodology considers these challenges. Another early work by Korolova et al.~\cite{korolova2010privacy} presents results using microtargeting to expose privacy leakage on Facebook. 

Other recent studies have focused on economics of display advertising~\cite{Gill2013}, characterizing mobile advertising~\cite{vallina2012breaking} and helping users to get control of their personal data and traffic in mobile networks \cite{rao2012meddle}, designing large scale targeting platforms~\cite{chen2009large} or investigating the effectiveness of behavioural targeting \cite{yan2009much}. Moreover, L\'ecuyer et al. \cite{lecuyer2014xray} developed a service-agnostic tool to establish correlation between input data (e.g., users actions) and resulting personalized output (e.g., ads). The solution is based on the application of the differential correlation principle on the input and output of several shadow accounts that generate a differentially distinct set of inputs.

Our work is different in focus to this previous literature since we are primarily concerned with OBA in display advertising, with the intention of understanding the collection and use of sensitive personal information at a large scale. To the best of the authors knowledge, only a couple of previous works analyze the presence of OBA advertising using a measurement driven methodology. Liu et al.~\cite{liu2013adreveal} study behavioural advertisement using complex end-user profiles with hundreds of interests  (instead of personas with specific interests) generated from an AOL dataset including users online search history. The extracted profiles from passive measurements are rather complex (capturing multiple interests and types), and are thus, rather inappropriate for establishing causality between specific end-user interests and the observed ads. Our approach is active rather than passive, and thus allows us to derive an exact profile of the interest that we want to capture. Furthermore, the authors collapse all types of targeted advertising (demographic, geographic and OBA), excepting re-targeting, whereas we focus on OBA due to its higher sensitivity from a privacy perspective. Barford et al.~\cite{barford2014adscape} present a large-scale characterisation study of the advertisement landscape. As part of this study the authors look at different aspects such as the new ads arrival rate, the popularity of advertisers, the importance of websites or the distribution of the number of ads and advertisers per website. The authors examine OBA very briefly. They trained personas but as they acknowledge their created profiles present a significant contamination including unrelated interests to the persona. Our methodology carefully addresses this issue.  Moreover, these previous works check only a small point of the entire spectrum of definitions, metrics, sources, filters, \etc. For instance, they rely on Google ads services to build their methodologies which reduces the generality of their results. Our work has taken a much broader look on OBA including both the methodology, the results, and the derived conclusions.
Finally, to the best of the authors knowledge, ours is the first work reporting results about the performance of the used methodology, the extent to which OBA is used in different geographical regions  and the utilization of DNT across the web.

%% file: conclusion.tex
\section{Conclusions}
\label{sec:conclusion}
This paper presents a methodology to identify and quantify the presence of OBA in online advertising. We have implemented the methodology into a scalable system and run experiments covering a large part of the entire spectrum of definitions, metrics, sources, filters, etc that allows us to derive conclusions whose generality is guaranteed. In particular, our results reveal that OBA is a technique commonly used in online advertising. Moreover, our analysis using more than 50 trained personas suggests that the volume of OBA ads received by a user varies depending on the economical value associated to the behaviour/interests of the user. More importantly, our experiments reveal that the online advertising market targets behavioural traits associated to sensitive topics (health, politics or sexuality) despite the existing legislation against it, for instance, in Europe. Finally, our analysis indicates that there is no significant geographical bias in the application of OBA and that do-not-track seems to not be enforced by publishers and aggregators and thus it does not affect OBA. These essential findings pave a solid ground to continue the research in this area and improve our still vague knowledge on the intrinsic aspects of the online advertising ecosystem.